\documentstyle[12pt,epsf,epsfig,wrapfig]{article}
\def\Journal#1#2#3#4{{#1} {\bf #2}, #3 (#4)}

\def\NIMA{{\em Nucl. Instrum. Methods} A}
\def\NPB{{\em Nucl. Phys.} B}
\def\PLB{{\em Phys. Lett.}  B}
\def\PRL{\em Phys. Rev. Lett.}
\def\PRD{{\em Phys. Rev.} D}

\def\EPJC{{\em Eur. Phys. J.} C}

\begin{document}
\begin{center}
{\bfseries SPIN EFFECTS IN LARGE RAPIDITY NEUTRAL PION PRODUCTION
AT STAR} \vskip 5mm \underline{D.A. Morozov} $^{1 \dag}$ for the
STAR collaboration \vskip 5mm
{\small (1) {\it Institute for High Energy Physics, Protvino, Russia}\\
$\dag$
{\it E-mail: Dmitry.Morozov@ihep.ru }}
\end{center}
\vskip 5mm

\begin{abstract}
Measurements by the STAR collaboration of neutral pion production at
large Feynman $x$ ($x_F$) in the first polarized proton collisions at
$\sqrt{s}=200$ GeV were reported previously.  During the following two
runs additional statistics were acquired with an improved forward
calorimeter for the $\pi^0$ cross-section and analyzing power measurements.
First data from $pp$ collisions at $\sqrt{s}=410$ GeV were taken during
the RHIC run that ended in June, 2005.

The cross section was measured at $\eta=$3.3, 3.8 and 4.0 and was found
to be consistent with next-to-leading order perturbative QCD calculations.

The analyzing power was found to be zero at negative $x_F$ and at positive
$x_F$ up to $~$0.3, then increased with increasing $x_F$. This behavior
can be described by phenomenological models including the Sivers effect,
the Collins effect or higher twist contributions in initial and final states.

Results for the analyzing power at $\eta=$3.7 and 4.0 from all data
acquired at $\sqrt{s}=200$ GeV and the status of the analysis of
the $\sqrt{s}=410$ GeV data will be presented. Future upgrade plans
and status will also be discussed.
\end{abstract}
\vskip 8mm

\section*{Introduction}
\label{sec:intro}
At present Quantum Chromodynamics (QCD) can notexplain the origin of
significant transverse single spin asymmetry($A_N$) in partonic
interactions. Collinear factorized perturbativeQCD (pQCD)
calculations at leading twist predict these analyzingpowers to be
entirely negligible, due to chirality in the theory. However,
experimental data \cite{E704,proza,E925} shows that $A_N$ for
inclusive particle production is on the order of $10\%$ independent
of the center of mass energy ($\sqrt{s}$). To improvethe situation
theorists develop several models in a generalized version of the
QCD factorization scheme, which allows for intrinsic transverse
motion of partons inside hadrons, and of hadrons relatively to
fragmenting partons. This adds new possibilities of spin effects,
absent for collinear configurations. Sivers \cite{Sivers} proposed
as a source of spin effects to be a flavor dependent correlation between
the proton spin (${\bf S_p}$),momentum (${\bf P_p}$) and transverse
momentum (${\bf k^\perp}$) of the unpolarized partons inside the proton.
This results in the new polarized parton distribution function:
\begin{equation}
\label{eq:sfcn}
f_{q/p^{\uparrow}}(x,{\bf k^{\perp}_q};{\bf S_p})=
\hat{f}_{q/p}(x,k^{\perp}_q)+
\frac{1}{2}\Delta^{N}f_{q/p^{\uparrow}}(x,k^{\perp}_q)
\frac{{\bf S_p}\cdot ({\bf P_p}\times{\bf k^{\perp}_q})}
{|{\bf S_p}||{\bf P_p}||{\bf k^{\perp}_q}|},
\end{equation}
where $\hat{f}_{q/p}(x,k^{\perp}_q)$ - unpolarized distribution function,
$\Delta^{N}f_{q/p^{\uparrow}}(x,k^{\perp}_q)$ - Sivers function and $x$
is the Bjorken scaling variable. Also significant $A_N$ could be produced
by the correlation between the quark spin (${\bf s_q}$), momentum
(${\bf p_q}$) and transverse momentum (${\bf k^\perp}$) of the pion in
the final state. Such an approach has been introduced by Collins
\cite{Collins}. Then the fragmentation
function of transversely polarized quark $q$ takes the form:
\begin{equation}
\label{eq:cfcn}
D_{\pi/q^{\uparrow}}(z,{\bf k^{\perp}_\pi};{\bf s_q})=
\hat{D}_{\pi/q}(z,k^{\perp}_\pi)+
\frac{1}{2}\Delta^{N}D_{\pi/q^{\uparrow}}(z,k^{\perp}_\pi)
\frac{{\bf s_q}\cdot ({\bf p_q}\times{\bf k^{\perp}_\pi})}
{|{\bf p_q}\times{\bf k^{\perp}_\pi}|},
\end{equation}
where $\hat{D}_{\pi/q}(z,k^{\perp}_\pi)$ - unpolarized fragmentation
function, $\Delta^{N}D_{\pi/q^{\uparrow}}(z,k^{\perp}_\pi)$ -Collins
function and $z$ is longitudinal component of pion momentum. Along with
Collins and Sivers mechanisms there are higher twist effects in either
initial \cite{QuiSterman} or final \cite{Koike} state which may cause
the observed analyzing powers.

The Relativistic Heavy Ion Collider (RHIC) at Brookhaven National Lab
(BNL) provides collisions of polarized protons at the energy of
$\sqrt{s}=200$ GeV since 2002. At 2005 first 410 GeV collisions
of polarized protons have been detected. The Solenoidal Tracker at
RHIC (STAR \cite{STAR}) consists mainly of a large volume TPC,
Forward TPC, Beam Beam Counters (BBC), Endcap Electromagnetic
Calorimeter (EEMC), Barrel Electromagnetic Calorimeter (BEMC) and
Forward Pion Detector (FPD). This contribution will focus on results
from FPD and BBC, which located in the very forward region of STAR
coverage. BBC are segmented scintillator detectors surrounding the
beam pipe. It provides the minimum bias trigger, absolute luminosity
and relative luminosity for our experiment. In addition BBC
coincidences are used to suppress beam gas background.
FPD is a set of eight calorimeters of lead glass cells
with size of 3.8 cm$\times3.8$ cm$\times45$ cm.
It provides triggering and reconstruction of neutral pions.
Four of them are left-right detectors and $7\times7$ arrays cells.
Four others are top-bottom $5\times5$ arrays and are useful
for systematics studies.

\section{Single Spin Asymmetry at STAR/FPD}
\label{sec:an}
By definition single spin asymmetry is:
$A_N=\frac{1}{P_{Beam}}\frac{d\sigma^\uparrow-d\sigma^\downarrow}
{d\sigma^\uparrow+d\sigma^\downarrow}$,
where $P_{Beam}$ - polarization of transversely polarized beam,
$d\sigma^{\uparrow(\downarrow)}$ - differential cross section of
$\pi^0$ when incoming proton has spin up (down). One can measure
$A_N$ by two different ways. First with the use of single arm calorimeter:
$A_N=\frac{1}{P_{Beam}}\frac{N^\uparrow-RN^\downarrow}
{N^\uparrow+RN^\downarrow}$, where $N^{\uparrow(\downarrow)}$ - the
number of pions detected when the polarization of the beam
is oriented up(down) and $R=\frac{L^\uparrow}{L^\downarrow}$
is the spin dependent relative luminosity measured by BBC.
Second with the use of two arms calorimeter (''cross-ratio'' method):
$A_N=\frac{1}{P_{Beam}}
\frac{\sqrt{N^{\uparrow}_LN^{\downarrow}_R}-\sqrt{N^{\uparrow}_RN^
{\downarrow}_L}}{\sqrt{N^{\uparrow}_LN^{\downarrow}_R}+
\sqrt{N^{\uparrow}_R N^{\downarrow}_L}}$,
where$N^{\uparrow}_{L(R)}$ - number of pions detected by the left
(right) calorimeter while the beam has spin up and
$N^{\downarrow}_{L(R)}$ - number of pions detected by the left
(right) calorimeter while the beam has spin down. In this method
one does not need the relative luminosity. The asymmetries from
these two measurements were found to be consistent.
Positive (negative) $x_F$ is defined when the pion is observed with
the same (opposite) longitudinal momentum as the polarized beam.
Positive $A_N$ is defined as more $\pi^0$ going left of the upward
polarized beam. In the 2002 proton run $0.15$ pb$^{-1}$ of
integrated luminosity was collected for transversely polarized
proton collisions at $\sqrt{s}=200$ GeV at an average polarization of
$16$$\% $. In 2003 run the integrated luminosity and average polarization
have been increased to $0.5$ pb$^{-1}$ and $27$$\% $ respectively.
In 2005(run 5) we collected 0.4 pb$^{-1}$ of $\sqrt{s}=200$ GeV data
with polarization of $45\%$. The polarization was measured by pC
CNIpolarimeter \cite{CNI}. All mentioned values of polarization
arebased on the online results from CNI polarimeter.

\begin{figure}
\begin{center}
\epsfig{figure=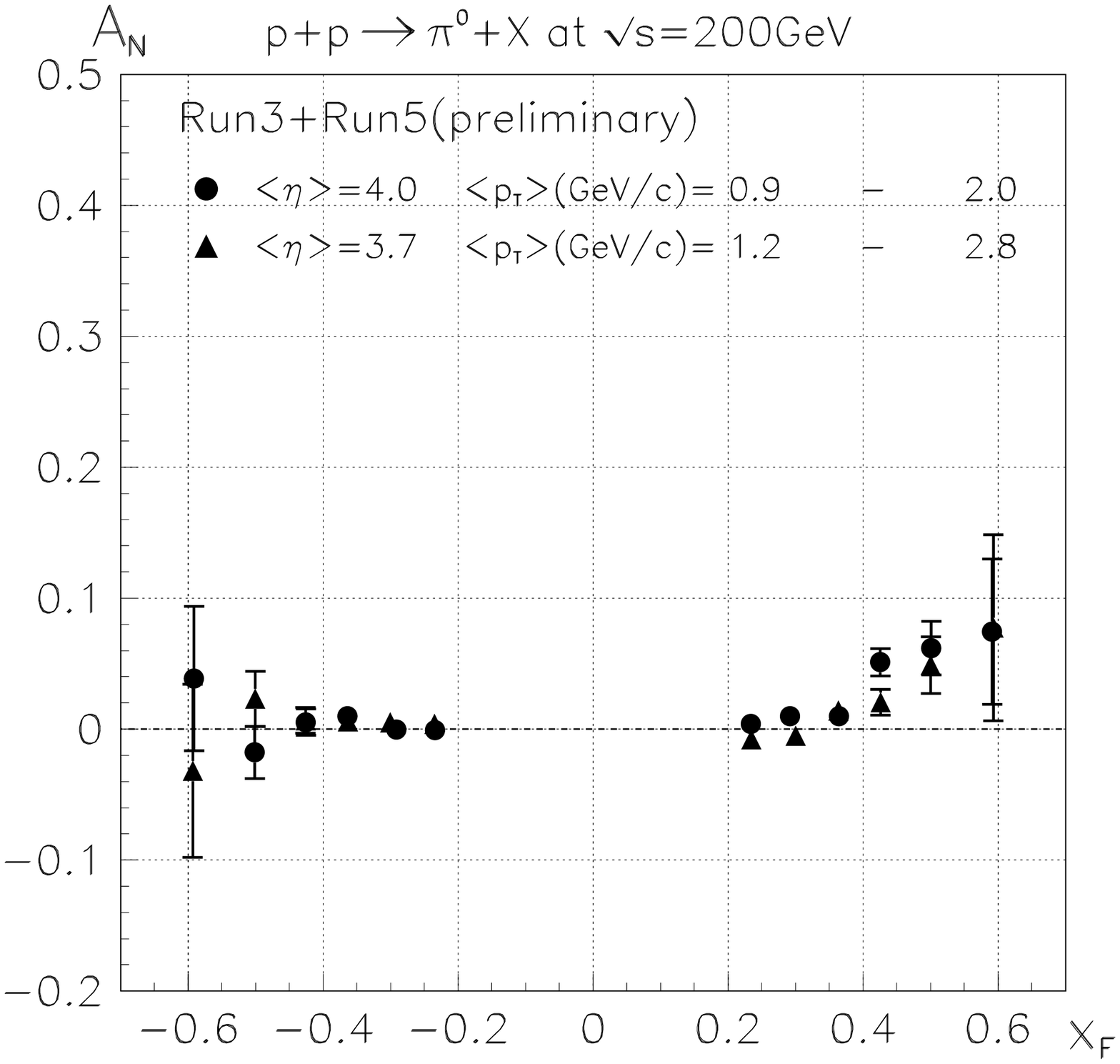,width=6.0cm}
\epsfig{figure=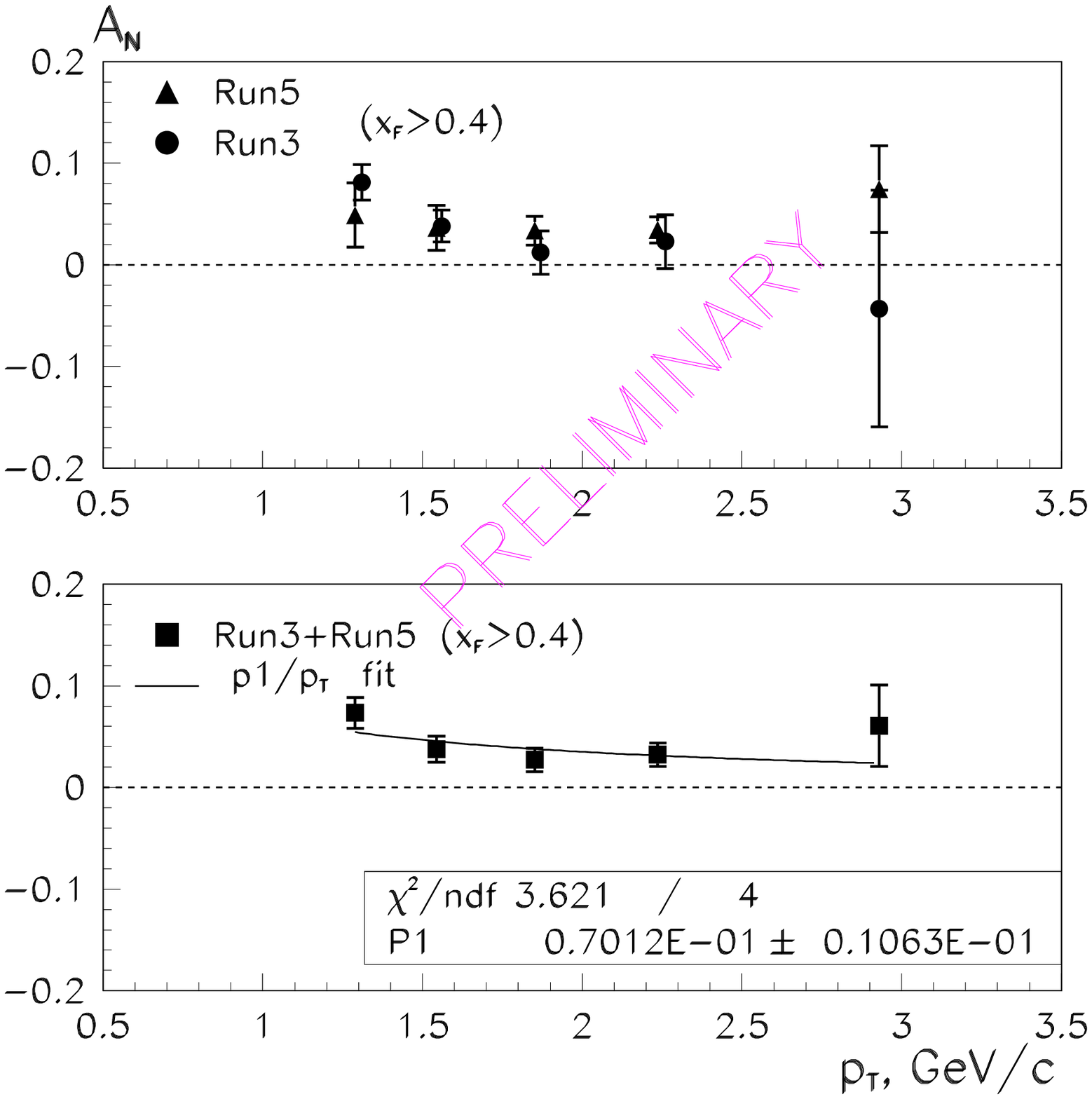,width=5.8cm}
\caption{\label{morozov_da_fig2} Left: Analyzing power of $\pi^0$ in
$p^{\uparrow}p\to\pi^0X$ reaction in run3 and run5 as a function of
$x_F$. Right-top: Analyzing power in run3 and run5 as a function of
$p_T$ at $x_F>0.4$. Right-bottom: Analyzing power as a function of
$p_T$ at $x_F>0.4$, combined data from run3 and run5.}
\end{center}
\end{figure}

Earlier result from published 2002 data for $<\eta>=3.8 $\cite{fpdprl}
consistent with measurements at lower $\sqrt{s}=20$GeV (E704 experiment)
and increasing with $x_F$. It also can be described by all theoretical
predictions mentioned above due to statistical uncertainties. Preliminary
results from run 2003 at $<\eta>=4.1$ were reported earlier \cite{Ogawa}.
The analyzing power for positive $x_F$ at $<\eta>=4.1$ is consistent with
zero up to $x_F\sim0.35$, then increases with increasing $x_F$. The
first measurement of $A_N$ at negative $x_F$ has been done, and is
found to be zero. Negative $x_F$ results may give an upper limit on the
gluon Sivers function \cite{Anselmino}. Using the same analysis as
in run3 we extracted analyzing power from run5. Preliminary results
from run5 data are shown at the Fig. \ref{morozov_da_fig2}.
Left plot represents $A_N(x_F)$ dependencies from run3 and run5
data. We combined data from two runs since the results are
consistent. There are two sets of points on the plot - for $<\eta>=3.7$ and
for $<\eta>=4.0$. These numbers reflects two different working
positions of the calorimeters -- closer and father to the beam
respectively. $A_N$ is nonzero at $x_F>0.4$ and zero for negative
$x_F$. The work on systematic errors calculations and single arm
results (for consistency check with ''cross-ratio'' method) is in
progress. On the right plot of Fig. \ref{morozov_da_fig2} one can
see $p_T$ dependence of analyzing power for $x_F>0.4$. Top plot
shows $A_N(p_T)$ for run3 and run5 data separately. In the bottom
plot we combined data from two runs. There is an evidence that
analyzing power at $x_F>0.4$ decreases with increasing $p_T$. To
interpret and compare these data with theory we need to add
systematic errors and single arm data.

\section{Differential cross sections for forward $\pi^0$-Production}
\label{sec:cros}
\begin{figure}[!h]
\begin{center}
\epsfig{figure=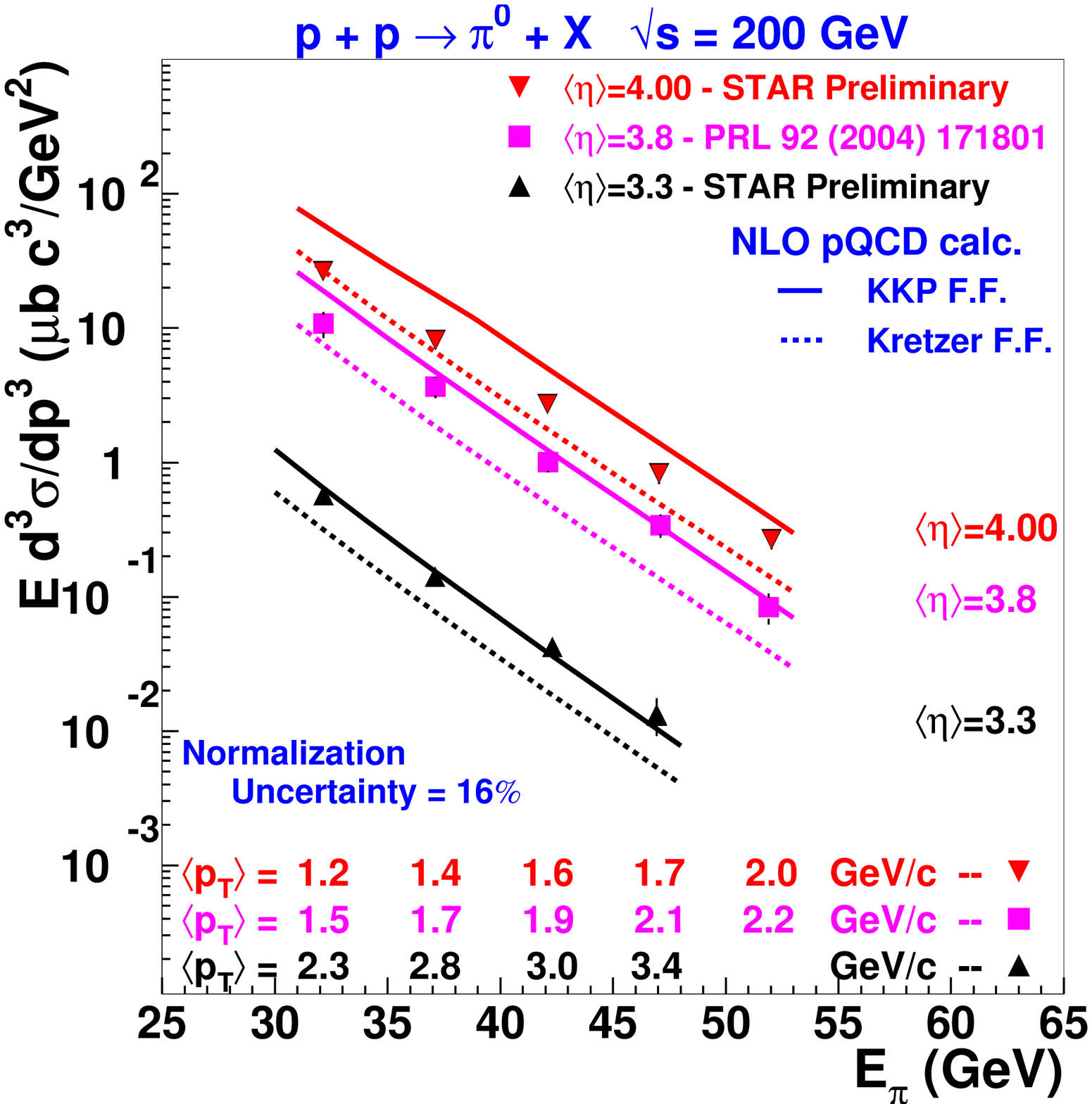,width=6.0cm}
\epsfig{figure=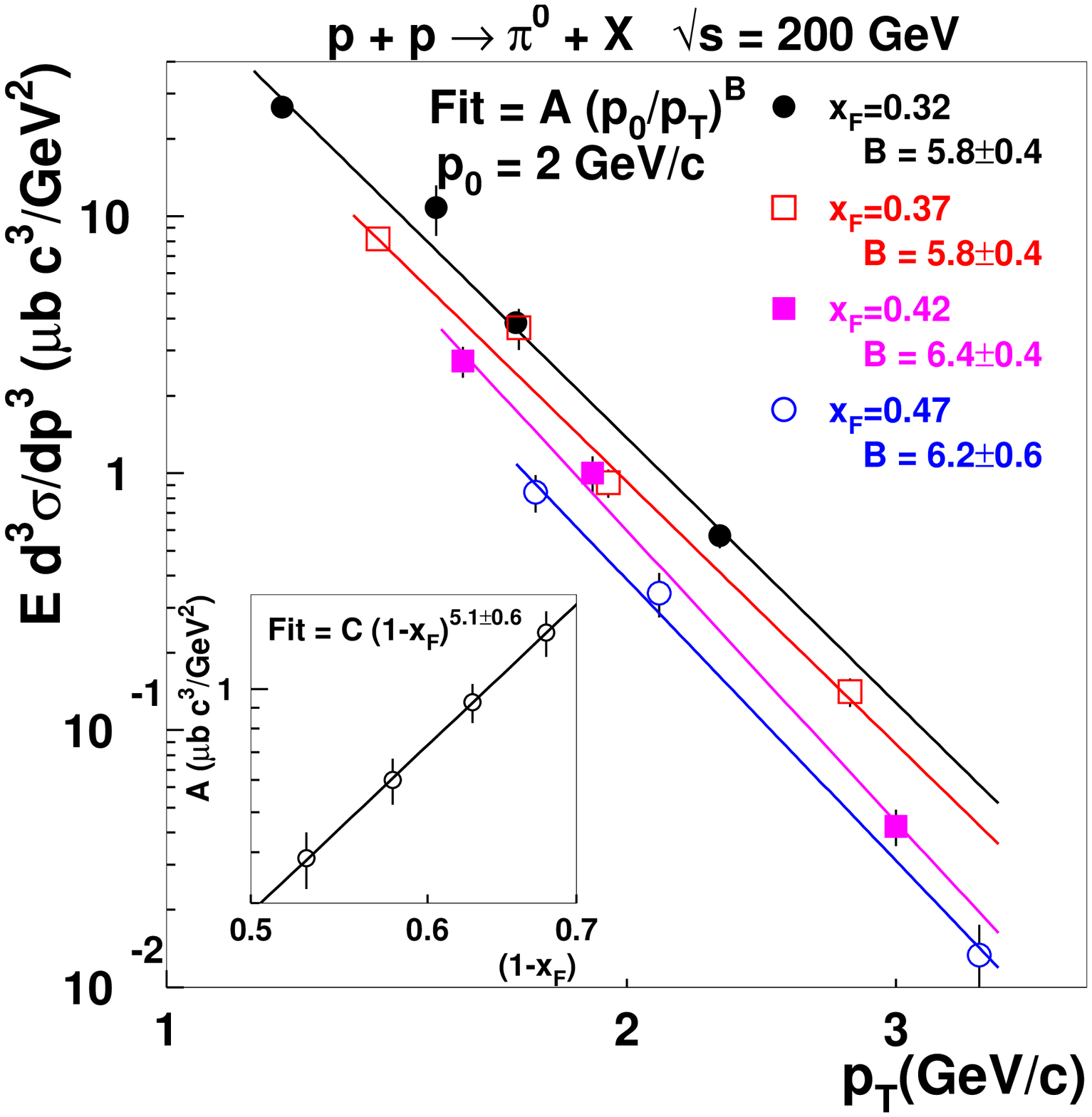,width=5.8cm}
\caption{\label{morozov_da_fig3} Left: invariant cross section for
$\pi^0$ produced in $pp$ collisions at $\sqrt{s}=200$ GeV versus
pion energy ($E_\pi$) at average pseudorapidities ($<\eta>$) $3.3$,
$3.8$ and $4.0$. The error bars are point-to-point systematic and
statistical errors added in quadrature. Right: invariant cross
section as a function of $p_T$ at fixed $x_F$ (outer) and as a
function of $(1-x_F)$ at fixed $p_T=2$ GeV/c (inner). Lines are fits
by the functions showed in the plot.}
\end{center}
\end{figure}

The inclusive differential cross section for
$\pi^0$ production for $30<E_\pi<55$ GeV at $<\eta>=3.8$ was
previously published \cite{fpdprl}. The result at $<\eta>=3.3$ in
2002 run also have been extracted. In 2003 run new calorimeters and
readout electronics have been installed to allow measurements of the
differential cross section at $<\eta>=4.0$. The results are shown in
Fig. \ref{morozov_da_fig3}.

On the left plot the cross sections are shown versus pion energy
and are compared with NLO pQCD calculations evaluated at $\eta=3.3$,
$3.8$ and $4.0$. Two sets of fragmentation functions are used. The
model calculations are consistent with the data in contrast to the
data at lower $\sqrt{s}$ (NLO pQCD calculations at $\sqrt{s}=20$ GeV
underpredict measured cross sections \cite{BorSof}). As $\eta$
increases, systematics regarding the comparison with NLO pQCD
calculations begin to emerge. The data at low $p_T$ are more
consistent with the Kretzer set of fragmentation functions. Similar
trend was observed at mid-rapidity \cite{phenixcs}. On the right
plot the data is represented as in earlier experiments \cite{ISR}.
The outer picture shows cross section as a function of $p_T$ at
fixed $x_F$. The inner one -- cross section as a function of
$(1-x_F)$ at fixed $p_T=2$ GeV/c. Invariant cross section falls with
$p_T$ at fixed $x_F$ with exponent (value $\sim6$) independently on
$x_F$. Data also show exponential dependence on $x_F$ with fixed
$p_T=2$ GeV/c. The value of the fitted exponent ($\sim5$) may be
sensitive to the interplay between hard and soft scattering
processes. One note should be stated regarding systematics of this
separated $x_F$ and $p_T$ dependencies. Data were accumulated in
different conditions in different running years: with different
calorimeters, with different readout electronics, taken in different
kinematical regions.

\section{Plans for the near-term future}
\label{sec:plans}
\begin{wrapfigure}[16]{r}{6cm}
\mbox{\epsfig{figure=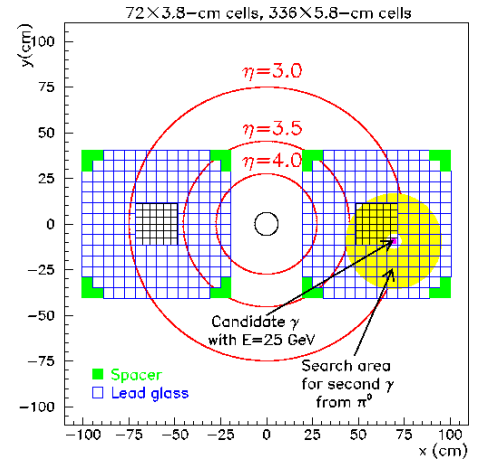,width=6cm,height=6cm}}
\caption{The FPD++ layout with isolation cut for single photon
events. Circles represent different values of $\eta=$3.0, 3.5, 4.0.}
\label{morozov_da_fig5}
\end{wrapfigure}

From the run5 data at $\sqrt{s}=410$ GeV we canextract unpolarized
differential cross section as we did for
$\sqrt{s}=200$ GeV. It will allow us to separate
$x_T=\frac{2p_T}{\sqrt{s}}$ and $p_T$ dependence of cross section:
$E\frac{d^3\sigma}{dp^2}\propto\left(\frac{\sqrt{s}}{2p_T}\right)^A
\left(\frac{1}{p_T}\right)^B(1-x_F)^N,$
where $N\approx 5$ and $A+B\approx 6$ (see Fig.\ref{morozov_da_fig3} right).
\begin{wrapfigure}{l}{6cm}
\mbox{\epsfig{figure=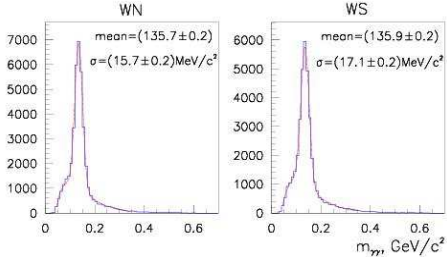,width=6cm,height=3.6cm}}
\caption{Di-photon invariant mass distributions for WS and WN
modules of FPD from run5 at $\sqrt{s}=410$ GeV.}
\label{morozov_da_fig4}
\end{wrapfigure}
This work is in progress. As an example of reconstruction of $\pi^0$
mesons at $\sqrt{s}=410$ GeV, the invariant mass distributions for
two different FPD modules are shown at Fig. \ref{morozov_da_fig4}.
The resolution in di-photon invariant mass distribution is better
than 20 MeV. Existing FPD detectors well suited to large rapidity
inclusive $\pi^0$ reconstruction.
The next step will be an implementation of interim forward
calorimeter FPD++ for RHIC run6 designed for single $\gamma$
detection (schematic view is shown at Fig. \ref{morozov_da_fig5}).
FPD++ will substitute west side of FPD detector. By design it
represents as two Lead glass counters matrices. Each consists of
14$\times$14 of 5.8$\times$5.8 cm$^2$ cells with a space 4$\times$4
in the middle for 6$\times$6 smaller cells (3.8$\times$3.8 cm$^2$).
FPD++ detectors will cover a broader range of $\Delta\eta$ and
$\Delta\phi$.

We have recently proposed to assemble a Forward Meson Spectrometer
(FMS) \cite{FMS}. The FMS will cover the range $2.5<\eta<4.0$ and
give STAR nearly hermetic electromagnetic coverage in the range $-1<\eta<4$.
With the addition of the FMS, which has more than 25 times
larger area coverage than the FPD, we will be able to achieve at
least three physics objectives: a measurement of the gluon density
distributions in gold nuclei for $0.001 < x < 0.1$; characterization
of correlated pion cross sections as a function of $p_T$ to search
for the onset of gluon saturation effects associated with
macroscopic gluon fields; measurements with transversely polarized
protons that are expected to resolve the origin of the large
transverse spin asymmetries in $p_{\uparrow}\ + p \to \pi^0+X$
reactions for forward $\pi^0$ production. The expected completion of
FMS is by October 2006.

\section*{Summary}
\label{sec:sum}
Large spin effects have been observed at forward
$\pi^0$ production in polarized $pp$ collisions at energy
$\sqrt{s}=200$ GeV at STAR FPD. The single spin asymmetry for
positive $x_F$ is consistent with zero up to $x_F=0.35$, then
increases with increasing $x_F$. The asymmetry is found to be zero
for negative $x_F$. First result for the $A_N(p_T)$ was obtained
using combined statistics from run3 and run5. There is an evidence
that the $\pi^0$ analyzing power at $x_F>0.4$ decreases with
increasing $p_T$. The inclusive differential cross section for
forward $\pi^0$ production at $\sqrt{s}=200$ GeV is consistent with
NLO pQCD calculations in contrast to what was observed at lower
energy. First try to map the cross section in $x_F-p_T$ plane was
performed. The near-future plans are planning enhancement of forward
calorimeter for RHIC run6 FPD++ for $\gamma/\pi^0$ separation. The
longer term future is upgrade in forward calorimetry in STAR (FMS)
to probe low-x gluon densities and establish dynamical origin of
$A_N$ (complete upgrade by October 2006).


\end{document}